\documentclass[preprint,12pt]{elsart}
\usepackage{latexsym}
\usepackage{amssymb}
\usepackage{graphicx}
\usepackage{epsfig}
\usepackage{wasysym}
\usepackage{esint}
\usepackage{subfigure}
\usepackage[font=small]{caption}
\usepackage{hyperref}

\newcommand{\beq}[0]{\begin{equation}}
\newcommand{\eeq}[0]{\end{equation}}

\newcommand{\bit}{\begin{itemize}}
\newcommand{\eit}{\end{itemize}}
\newcommand{\bea}{\begin{eqnarray*}}
\newcommand{\eea}{\end{eqnarray*}}
\newcommand{\beanmb}{\begin{eqnarray}}
\newcommand{\eeanmb}{\end{eqnarray}}
\newcommand{\ben}{\begin{equation}}
\newcommand{\een}{\end{equation}}

\newcommand{\R}{\mathbb{R}}
\renewcommand{\etal}{{\it et al.}}

\begin{document}
\newlength{\caheight}
\setlength{\caheight}{12pt}
\multiply\caheight by 7
\newlength{\secondpar}
\setlength{\secondpar}{\hsize}
\divide\secondpar by 3
\newlength{\firstpar}
\setlength{\firstpar}{\secondpar}
\multiply\firstpar by 2

\begin{frontmatter}
\vskip 48pt
\title{$H\to \gamma \gamma:$ a Comment on the 
Indeterminacy\\ of Non-Gauge-Invariant Integrals}
\author[1]{F.~Piccinini},
\author[2]{A.~Pilloni} and 
\author[2,3]{A.D.~Polosa}
\address[1]{Istituto Nazionale di Fisica Nucleare, Sezione di Pavia\\
Via A. Bassi 6, I-27100, Pavia, Italy}
\address[2]{Dipartimento di Fisica, Sapienza Universit\`a di Roma\\
P.le A. Moro 2, I-00185, Roma, Italy}
\address[3]{Istituto Nazionale di Fisica Nucleare, Sezione di Roma\\
P.le A. Moro 2, I-00185, Roma, Italy}
\begin{abstract}
We reanalyze the recent computation of the amplitude of the Higgs boson decay 
into two photons presented  by 
Gastmans {\etal}~\cite{Gastmans:2011ks,Gastmans:2011wh}. 
The reasons for which  this result cannot be the correct one have been  discussed 
in some recent papers. We address here the general issue of the indeterminacy of 
integrals  with four-dimensional gauge-breaking regulators 
and  to which extent it might eventually be solved by  imposing 
physical constraints. 
Imposing gauge invariance as the last step upon  $R_{\xi}$-gauge calculations 
with four-dimensional gauge-breaking regulators,
allows indeed to recover the well  known  
$H \to \gamma\gamma$ result.
However we show  that in the particular case of  the unitary gauge, 
the indeterminacy cannot be tackled in  this same 
way. The combination of unitary gauge with a cutoff regularization 
scheme turns out to be non-predictive. 
\end{abstract}

\end{frontmatter}
\newpage
\section{Introduction}
Recently some attention has been brought back to
the $W$-loop contribution  in the calculation of the $H\to \gamma\gamma$ amplitude
because of a result presented by Gastmans {\etal}~\cite{Gastmans:2011ks,Gastmans:2011wh} 
turning out to be at odds with the renowned one of Refs.~\cite{Ellis:1975ap,Shifman:1979eb}.
It goes without saying that, if correct, the result in Refs.~\cite{Gastmans:2011ks,Gastmans:2011wh}  would have had  relevant consequences for the ongoing Higgs boson searches at the LHC.
 
Starting from the observation that the full amplitude 
$H\rightarrow\gamma\gamma$ is free from ultraviolet and infrared 
singularities, Gastmans {\etal} performed their calculation in 
four dimensions with no regulators and used the unitary gauge 
to consider only the physical degrees of freedom.
A gauge invariant amplitude is obtained with  
the `Dyson subtraction'~\cite{Dyson:1949bp,Dyson:1949ha}, 
leading to
\begin{equation}
{\cal M}=\frac{e^2g}{(4\pi)^2m_W}\left[3\tau+3\tau(2-\tau)f(\tau)\right]
(k_1\!\cdot k_2\,g^{\mu\nu}-k_2^\mu k_1^\nu)\epsilon_\mu(k_1) \epsilon_\nu(k_2)
\label{eq:amplitudewrong}
\end{equation}
where $\tau = \frac{4 m_W^2}{m_H^2}$ and
\begin{equation}
f(\tau)=\left\{
\begin{array}{ll}
\arcsin^2(\tau^{-\frac{1}{2}}) &\quad \mbox{for}\quad \tau\geq 1\\
-\frac{1}{4}\left[\ln\frac{1+\sqrt{1-\tau}}{1-\sqrt{1-\tau}}-i\pi\right]^2&
\quad \mbox{for}\quad \tau<1
\end{array}\right.\;
\end{equation}
This amplitude, which happens to vanish in the  $m_W/m_H \to 0$ limit 
(contrary to the standard one), would imply a reduction of the decay width 
$\Gamma(H\rightarrow\gamma\gamma)$ by about $50$\% for $m_H \simeq 120$ GeV, 
with respect to what found in Refs.~\cite{Ellis:1975ap,Shifman:1979eb}. 

The standard $H\rightarrow\gamma\gamma$ amplitude
was computed in `t~Hooft-Feynman gauge with dimensional 
regularization~\cite{Ellis:1975ap}, with background field 
methods~\cite{Shifman:1979eb}, and in unitary gauge 
with renormalization group analysis~\cite{Shifman:1979eb}. It reads as
\begin{equation}
{\cal M}=\frac{e^2g}{(4\pi)^2m_W}\left[2 + 3\tau+3\tau(2-\tau)f(\tau)\right]
(k_1\!\cdot k_2\,g^{\mu\nu}-k_2^\mu k_1^\nu)\epsilon_\mu(k_1) \epsilon_\nu(k_2)
\label{eq:amplitudecorrect}
\end{equation}
Gastmans {\etal} casted some  doubts on the reliability of using 
dimensional regularization and prefer their result adding  the  motivation that  
it would respect some $m_W/m_H \to 0$   `decoupling limit', which has indeed no 
reason to hold true as explained in Refs.~\cite{Shifman:2011ri,Huang:2011yf,Marciano:2011,Jegerlehner:2011jm}
in the framework of equivalence theorem \cite{equivalenceth}.

The results of Refs.~\cite{Gastmans:2011ks,Gastmans:2011wh} 
have been criticized by a number of recent papers~\cite{Shifman:2011ri,Huang:2011yf,Marciano:2011,Jegerlehner:2011jm,Shao:2011,liang-czarnecki,Bursa:2011aa}. 

The criticism concerns the absence of regulators, 
leading to ambiguities in the intermediate steps of the calculation, 
the use of Dyson subtraction and the reference to the Appelquist-Carazzone 
theorem~\cite{Appelquist:1974tg} to justify the decoupling. 

The amplitude $H\to\gamma\gamma$ 
has been calculated in several ways, all confirming the 
old result of Refs.~\cite{Ellis:1975ap,Shifman:1979eb}, as follows: 
\begin{itemize}
\item  The authors of Ref.~\cite{Huang:2011yf} redo the calculation 
in four dimensions in the unitary gauge 
with a gauge-invariant regularization 
method (Pauli-Villars like)~\cite{lore}. They stress the importance 
of having set an explicit regularization scheme 
to control finite terms which guarantee gauge invariance through 
all intermediate steps of the calculation. 
The authors cross-check the calculation 
with an independent one in dimensional regularization and underscore 
that no renormalization condition should be applied in the presence 
of only finite terms. Thus they conclude that 
the calculation presented by Gastmans {\etal} must be wrong because 
it is finite and not gauge-invariant (essentially because 
of the lack of regularization of divergent integrals, finite terms, 
relevant for the gauge invariance of the final result, have been missed). 
\item  The authors of Ref.~\cite{Marciano:2011} perform the calculation 
in dimensional regularization  both in the unitary and in 
 $R_\xi$ gauges (the same as in Ref. \cite{Jegerlehner:2011jm}).
 It is  pointed out that without any regulator the coefficient of $g^{\mu \nu}$, arising upon 
four-dimensional symmetric integration in renormalizable gauges,
is  {\it an indeterminate form of the kind \mbox{$\infty-\infty$} } --
also responsible of the breaking of gauge invariance --
and that in the unitary gauge the same happens to the coefficient of 
$k_2^\mu k_1^\nu$.
With reference to Jackiw \cite{Jackiw:1999qq}, 
the authors stress the importance of symmetry requirements (in this case, 
gauge symmetry) to solve the  ambiguities in the finite terms occurring in loop calculations.
\item H.S. Shao {\etal} \cite{Shao:2011} perform the calculation 
using a four-dimensional momentum cutoff regularization 
both in `t~Hooft-Feynman and unitary gauges. Within the latter they obtain 
the same result of Refs.~\cite{Gastmans:2011ks,Gastmans:2011wh} starting with 
a particular routing of momenta. They  also get the terms to be added 
after a shift in the loop momentum: these contributions 
sum up to zero when  the momentum choice 
of Refs.~\cite{Gastmans:2011ks,Gastmans:2011wh} is adopted. 
Performing the calculation 
in the `t~Hooft-Feynman gauge, the authors recover the same gauge invariant 
result as the one obtained in 
dimensional regularization~\cite{Ellis:1975ap} 
by subtracting the countribution of all diagrams 
(evaluated at $k_1 = k_2 = 0$) except for the  $\phi^- \phi^- W^-$ loop, the one out of the three which is independent of $k_1,k_2$. 
The result is independent of the loop momentum choice 
because the divergences are only logarithmic. 
\item F.~Bursa {\etal}~\cite{Bursa:2011aa} 
perform the calculation of the $H \to \gamma \gamma$ 
decay amplitude using a (gauge invariant) 
spacetime lattice regulator and obtain very good numerical 
agreement with the decay amplitude 
evaluated  with dimensional regularization. 
\end{itemize}

Summarizing, in Ref.~\cite{Marciano:2011} it is highlighted  that the problem 
of Refs.~\cite{Gastmans:2011ks,Gastmans:2011wh} resides 
in the absence of regulators.
In Ref.~\cite{Shao:2011}  however it is shown that the use of cutoff regularization in unitary gauge leads to confirm the result by Gastmans {\etal}. Thus we might observe that, if there is a problem in the latter calculation, it is not in the lack of a regulator but rather in the combination of the unitary gauge  with the use of non-gauge invariant regulators. 
As the cutoff regularization has been widely used in the literature, 
we explore further its connection with the unitary gauge: 
in particular we attempt to get a deeper understanding of the 
result presented in Ref.~\cite{Shao:2011}. 

In Section~\ref{sec:2} we focus on the critical integrals at the core of the calculation of the $H\to\gamma\gamma$ amplitude. We stress, with some elementary examples, that if a cutoff regulator is chosen, the definition  of the integration boundaries is part of the regularization scheme itself.  In the paper by Jackiw~\cite{Jackiw:1999qq} it was clearly discussed how the critical integrals we have to deal 
with are indeterminate, as long as we do not use a regularization 
scheme which preserves the full symmetry of the theory.
Some two-dimensional examples are left in the Appendix~\ref{sec:appendix}
This would be enough to close the argument here. 
Yet  we show how the indeterminacy in cutoff regularization 
(Section~\ref{sec:3}) can  lead to the correct result 
if an appropriate finite renormalization condition is used. 

However in the case of the unitary gauge we will show this 
is not possible. 
In Section~\ref{sec:4} we extend the discussion to $R_\xi$ gauges. 
Our conclusions are left to Section~\ref{sec:5}.

\section{The Critical Integrals in the $H\to \gamma\gamma$ Amplitude}
\label{sec:2}
At the core of the problem of the $H\to \gamma\gamma$ amplitude calculation  there 
is the calculation of the integral
\begin{equation}
 I_{\mu\nu} = \int d^4 l \, 
\frac{g_{\mu\nu} l^2 - 4 l_\mu l_\nu}{\left(l^2 - M^2 + i\epsilon\right)^3} 
\label{eq:integral4D}
\end{equation}
where $M^2 = m_W^2 - x_1 x_2 m_H^2$, and $x_1,x_2$ are Feynman parameters.
According to  Gastmans {\etal}, performing the integral in four dimensions 
with symmetric boundaries~\footnote{Since the integral in Eq.~(\ref{eq:integral4D}) 
does not depend on any external momenta, 
for tensor invariance it must be $I_{\mu\nu} = I g_{\mu\nu}$; 
by saturating both sides with $g^{\mu\nu}$, 
we have $I_{\mu\nu} g^{\mu\nu} \rightarrow 4 I$, 
$l^2 g_{\mu\nu}\rightarrow 4 l^2$ and $l_\mu l_\nu\rightarrow l^2$, 
then we can solve with respect to $I$. We have the same result if we 
substitute $l_\mu l_\nu\rightarrow \frac14 l^2$.}, 
we can substitute $l_\mu l_\nu \rightarrow \frac14 l^2 g_{\mu\nu}$, 
leading to $I_{\mu \nu} = 0$.
Here we can appreciate the difference with respect 
to dimensional regularization (DREG), 
where it is found
\begin{equation}
I^\textup{DREG}_{\mu\nu}\left(n\right) = \int d^n l \, \frac{g_{\mu\nu} l^2 - 4 l_\mu l_\nu}{\left(l^2 - M^2 +i\epsilon\right)^3} = -i g_{\mu\nu} \frac{\pi^2}2 + O\left(n-4\right)
\end{equation}
Gastmans {\etal} conclude that 
$I^\textup{DREG}_{\mu\nu}\left(n\right)$ must have a {\it discountinuity in \mbox{$n=4$}}, thus mining the  foundations of the DREG technique stating that integrals are not analitic in $n$ dimensions.

Let us start from the  four-dimensional integral in Eq.~(\ref{eq:integral4D}).
After  Wick rotation~\footnote{Which amounts to $d^4 l=id_4 l$, $g_{\mu\nu}\to -\delta_{\mu\nu}$, $l^2 \to l_E^2$ and $l_\mu\to l_\mu^E$.} and rescaling $l\to l/M$, we get
\begin{equation}
 I_{\mu\nu} = i \int d_4 l \, 
\frac{\delta_{\mu\nu} l^2 - 4 l_\mu l_\nu}{\left(l^2 + 1\right)^3} 
\label{eq:integral4DWick}
\end{equation}
To simplify the discussion, let us focus on the case $\mu=1,\nu=1$
\begin{equation}
 I_{11} = i \int d_4 l \, \frac{l^2 - 4 l_1^2}{\left( l^2+1\right)^3} = i \int d_4 l\, F_{11}\left(l\right)
\label{eq:integral11}
\end{equation}
The integrand is not a summable function: it is not positive everywhere in the domain of integration, and $\int d_4 l\, \left| F_{11} \right| = \infty$, which means that the integral is not defined \emph{per se} - the value depends on how the boundary is chosen to behave at infinity. Let us therefore compute the value of the integral over different integration domains with different behaviors at infinity: we will observe how the integral may assume every finite value, and even diverge.

As a first example, let us consider a `spherical cutoff' in the sense described below. In polar coordinates, we write
\begin{eqnarray}
 I_{11} &=& i \int_0^\Lambda dl \, \frac{l^5}{\left(1 + l^2\right)^3} \, \int d\Omega_4 \, \left(1 - 4 \cos^2 \theta\right) \nonumber \\
&=& i 4 \pi \int_0^\Lambda dl \, \frac{l^5}{\left(1 + l^2\right)^3} \, \int_0^\pi d\theta \, \sin^2 \theta \left(1 - 4 \cos^2 \theta\right) = 0
\label{eq:integralspherical}
\end{eqnarray}
$\Lambda$ is a dimensionsless cutoff, being $l$ a dimensionsless 
integration variable. The angular part vanishes, 
so there are no problems with the logarithmic divergence of the radial part. 
Actually, every integration domain which has the  
$l_i \leftrightarrow \pm l_j$ symmetry, 
leads to an identically vanishing integral. 

\begin{figure}
\centering%
\includegraphics[width=7cm]{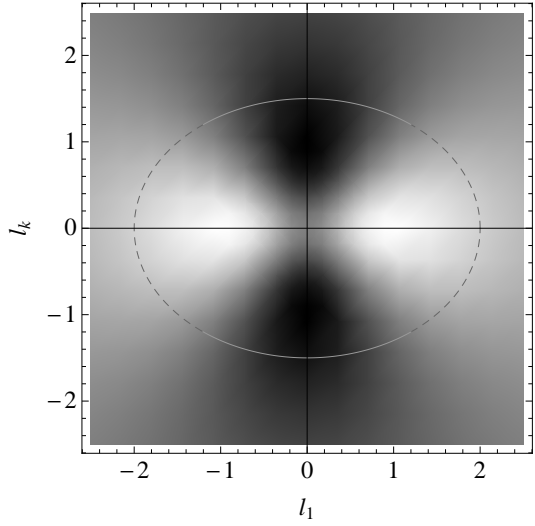}%
\caption{We integrate $F_{11} = \frac{l^2 - 4 l_1^2}{\left(1 + l^2\right)^3}$ over an elliptic domain. Because of cylindrical symmetry, the graphic is the same independently of $l_k =l_2,l_3,l_4$. The darker is the background, the larger is the $F_{11}$ value. The boundary is solid when $F_{11}>0$, dashed otherwise. Since we broke the spherical symmetry, the ellipse bounds a negative part which is larger than the positive one: the integral does not vanish.}%
\label{fig:ellipse}%
\end{figure}

As a second case we choose a non-symmetrical domain of integration. For example, let us integrate $F_{11}$ over the elliptical domain $\frac{l_1^2}{1+\epsilon} + l_2^2 + l_3^2 + l_4^2 \leq \Lambda^2$ (see \figurename{} \ref{fig:ellipse})
\begin{eqnarray}
 I_{11} &=& i 4 \pi \sqrt{1+\epsilon} \int_0^\Lambda dl \, l^5 \, \int d\theta \sin^2 \theta \, \frac{1- (4+3\epsilon) \cos^2 \theta}{\left(1 + l^2 + l^2 \epsilon \cos^2 \theta\right)^3 } \nonumber \\
 &=& i \pi^2 \frac{8 + 4\epsilon - \epsilon^2 -8 \sqrt{1+\epsilon}}{2 \epsilon^2} + O\left(\frac1{\Lambda^2}\right)
\label{eq:integralelliptical}
\end{eqnarray}
The integral  in Eq.~(\ref{eq:integralelliptical}) can assume different finite values as a function of  $\epsilon$. 

Choosing asymmetric boundaries, we even loose tensor invariance, obtaining a $4\times 4$ matrix of unrelated, indeterminate terms. This translates into the fact that $I_{\mu\nu}$ is no
longer proportional to $\delta_{\mu\nu}$ as it should be the case (see Eq~(\ref{eq:integral4DWick})).
We seek  an appropriate choice of  the boundaries for all the terms in the $I_{\mu\nu}$ matrix in such a way to recover a $\delta_{\mu\nu}$ structure. We can therefore compute the $I_{\mu\nu}$ entries by choosing the same asymmetric boundary on all diagonal terms, and generic symmetric boundaries for all off-diagonal terms. In this way, all diagonal terms will have the same indeterminate value $I$, whereas off-diagonal terms will vanish. We thus obtain $I_{\mu\nu} = I \delta_{\mu\nu}$, with $I$ being an indeterminate (even divergent) constant.

In the Appendix~\ref{sec:appendix} we give more details on this and we also consider the case of Schwinger regularization. We find that, also in this case, the value of the integral changes with the angular dependence of the cutoff function.
Schwinger regularization turns out to be particularly instructive as the indeterminacy of the integrals is not related to the shape of the integration domain but to the behavior at infinity of the integrand function.
This is a signal of the fact that, in order to tackle the indeterminacy of the critical integrals in this calculations, one exclusively needs to set  {\it physical constraints} as there is no mathematical prescription which can univocally determine them.

Gastmans {\etal} rely on their finite (equal to zero) result 
for the integral of the type of~Eq.~(\ref{eq:integral4D}), 
which follows from a {\it particular choice of the integration domain}.  
This also explains why the calculation of Ref.~\cite{Shao:2011}, 
in which a spherical cutoff is explicitely used, 
leads to the same result found by Gastmans {\etal} in unitary gauge: 
the choice of the integration domain in Ref.~\cite{Shao:2011} is the same as the one implicitely taken in Refs.~\cite{Gastmans:2011ks,Gastmans:2011wh}. 

The authors of Ref.~\cite{Marciano:2011}, on the other hand, 
underscore the fact that the 
integral~(\ref{eq:integral4D}) is an indeterminate $\infty-\infty$ form: it must be treated with some regularization scheme. Gauge-invariance in the final result  can be implemented either \emph{a priori} 
by choosing gauge-invariant regulators (like Pauli-Villars or DREG), 
or \emph{a posteriori} by applying an appropriate subtraction. 
We remark that in the latter case integrals are not well defined, 
and their  values must be considered indeterminate.

If we choose a sharp spherical cutoff we still get the Gastmans {\etal} 
result in unitary gauge as was first shown in Ref.~\cite{Shao:2011}. 
On the other hand, if we use renormalizable gauges in a cutoff scheme, 
we recover the standard result 
(`t~Hooft-Feynman gauge is used in Ref.~\cite{Shao:2011}).  

Does the fact that two different results are obtained using two different gauge choices mean that a cutoff scheme is not to be pursued at all? 
In the following we mean to show that the problem does not reside in the use of a cutoff scheme by itself but rather in figuring out that different ways (spherical, elliptical etc.) of implementing a cutoff scheme amount to different values of the integrals, {\it i.e.}, to indeterminate coefficients. We can actually use some cutoff scheme 
provided that there is a clear recipe on how to absorb the indeterminate coefficients arising in the calculation, also restoring gauge invariance. 
{\it We will show that such a recipe cannot be found in unitary gauge}. 
 
After this discussion one might therefore ask why cutoff renormalization  
works in several cases independently of 
the issue of the indeterminate constants discussed above.
 
For example, let us discuss the cutoff renormalization of vacuum polarization 
in QED. Since we are not protected by Ward identity, 
the integral presents a quadratic divergence
\begin{equation}
\Pi_2^{\mu\nu}\left(k\right) = -4i e_0^2 \int^\Lambda \frac{d^4 p}{\left(2\pi\right)^4} \frac{2 p^\mu p^\nu - g^{\mu\nu} p^2}{\left(p^2 - m^2\right)\left(\left(p - k\right)^2 - m^2\right)} \propto e_0^2 \Lambda^2 g^{\mu\nu} 
\end{equation}
The integral is divergent and sign-undefined; it is ill-defined as the one in Eq.~(\ref{eq:integral4D}). We could repeat the above considerations to show that we can lower the degree of divergence with an appropriate choice of boundary. Anyway we simply add a photon mass counterterm and impose that the quadratic term vanishes (i.e. that the photon is massless), in order to recover a good gauge invariant theory. The presence of the counterterm assures that the quadratic divergence disappears {\it whatever value the integral has}. In this case, we can forget about the indeterminacy itself, and calculate the integral with the boundary we prefer.
On the other hand, we cannot choose a boundary and claim that an integral like the one in Eq.~(\ref{eq:integral4D}) has a finite value and needs no counterterm (see Gastmans {\etal}); we must consider expressly the indeterminacy and add a counterterm to absorb it. 

\section{Considerations on power counting and gauge invariance}
\label{sec:3}
Jackiw \cite{Jackiw:1999qq} showed that indeterminacy 
can arise if we use regulators which have less symmetry than the theory, 
and it is not necessarily resolved when we restore the symmetry at the end 
of calculation. On the other hand, if the regulator maintains the symmetry, 
the result will be univocal. We can simply understand the case of 
gauge symmetry: gauge invariant regulators decrease the degree of divergence of 
the integral, making it finite and regulator independent. 
Indeed, by naive power counting, we know that the amplitude 
$H \rightarrow \gamma\gamma$ is logarithmically divergent in renormalizable gauges. 
In gauge invariant regularizations, we can group two momentum powers 
in the numerator to extract the gauge invariant 
factor $k_1 \!\cdot k_2\,g^{\mu\nu} - k_2^\mu k_1^\nu$, 
so that the amplitude becomes finite. In unitary gauge too, we expect the same finite amplitude after a not-straightforward cancellation of higher divergent terms.
However, this cannot be done in cutoff regularization where the Ward identity and gauge invariance are spoiled by the breaking of shift invariance. In the latter case 
the integral remains indeterminate or at worse logarithmically divergent.
The expectations by Gastmans {\etal} to get a finite amplitude which needs no regulator are disappointed by the choice of four-dimensional symmetric integration, which implicitly uses a spherical cutoff scheme, leading to the breaking of gauge invariance and to a divergent amplitude.

To subtract the divergence in the cutoff regularization scheme and, in general, all 
cutoff-dependent terms, we need some counterterms.
Breaking gauge symmetry, we have the most general lagrangian 
with all possible combinations of bare fields and bare couplings, 
even an \emph{ad hoc} counterterm of the form $\delta m_{0\!A}\, h_0 A_0^2$. 
This is what Dyson subtraction means: hide all divergent, 
cutoff-dependent, non-gauge-invariant terms into a counterterm, 
which would vanish in a gauge invariant regularization scheme.

We have computed $H \to \gamma\gamma$ amplitude in 
`t~Hooft-Feynman gauge without calculating divergent integrals; we find: 
\begin{eqnarray}
{\cal M}^{\mu\nu}_{\xi=1} &=& \frac{e^2g}{(4\pi)^2m_W} \Biggl[ -k_2^\mu k_1^\nu \left(2 + 3\tau + 3\tau\left(2-\tau\right)f\left(\tau\right)\right) \nonumber \\
&-& 2 m_H^2 \left(1 + \frac32 \tau\right)\!\int_0^1\!dx_1 \int_0^{1-x_1}\!\!dx_2 \int\!\frac{d^4l}{i\pi^2} \frac{g^{\mu\nu}l^2 - 4 l^\mu l^\nu}{\left(l^2 - 1 + 4x_1 x_2 \tau + i\epsilon\right)^3}  \nonumber \\
        &+& \frac12 m_H^2 g^{\mu\nu} \left( 1 + \frac{3}{2} \tau + 3 \tau\left(2 - \tau\right)  f\left(\tau\right) \right)\Biggr]
\label{eq:amplitudefeynman}
\end{eqnarray}

We remark that the first term (proportional to $k_2^\mu k_1^\nu$) 
contains only well-defined finite integrals; 
the second term is indeterminate 
(vanishing according to symmetric integration as 
in Refs.~\cite{Gastmans:2011ks,Gastmans:2011wh}). With the use of DREG, 
the second term would give 
$\frac{1}{2} m_H^2 g^{\mu\nu}\left(1 + \frac32 \tau\right)$, 
leading to the well known gauge-invariant expression. 
However, let us stay in the framework of gauge-breaking regularizations. 

In Ref.~\cite{Shao:2011} a modified version of Dyson subtraction 
is performed to recover gauge invariance.
One might even  wonder whether Dyson subtraction 
is allowed without divergent terms~\cite{Huang:2011yf}. As we have just shown, 
the integral in the second term is probably divergent and in any 
case cutoff-dependent, so we are allowed to add a counterterm and impose 
gauge invariance as a renormalization condition. 
In so doing we get the correct amplitude in Eq.~(\ref{eq:amplitudecorrect}). 
We would have the same expression by using symmetric integration: 
every value of the integral disappears into the counterterm. 
We are therefore lead to observe that  
the arbitrariness related to the choice of the boundary 
(or, in general, of the regulator) is solved by imposing gauge invariance.

Why Gastmans {\etal}'s amplitude is different from the standard one? 
As shown in Ref.~\cite{Gastmans:2011wh}, in unitary gauge we have 
another divergent integral
\begin{eqnarray}
A' &=& 2 \int_0^1 dx_1 \, \int_0^{1-x_1} dx_2 \, \int d^4 l \, \Biggl[
\left(k_2^\mu k_1^\nu - k_1 \cdot k_2 g^{\mu\nu} \right)l^2 
-2 k_1^\nu \left(k_2\cdot l\right) l^\mu \nonumber \\
&{}& -2 k_2^\mu \left(k_1\cdot l\right) l^\nu 
+ 2 \left(k_1\cdot k_2\right) l^\mu l^\nu 
+ 2 g^{\mu\nu} \left(k_1\cdot l\right)\left(k_2\cdot l\right) 
\Biggl] \nonumber \\
&{}& \cdot\frac{1}{\left(l^2 - m_W^2 + x_1 x_2 m_H^2 + i\epsilon \right)^3}
\label{eq:aprimo}
\end{eqnarray}

By symmetric integration the integral vanishes, whereas dimensional 
regularization leads to $A'_\textup{DREG}\left(n\right) = i \pi^2 \left(k_2^\mu k_1^\nu - k_1\cdot k_2 g^{\mu\nu}\right) + O\left(n - 4\right)$. 
The integral has the same indeterminate behavior of the former one: 
the value depends on the choice of the boundary. 
We can say that $A' = J k_2^\mu k_1^\nu + J' g^{\mu\nu}$, 
with $J$ and $J'$ indeterminate constants. 
While in the Eq.~(\ref{eq:amplitudefeynman}) the tensor $k_2^\mu k_1^\nu$ 
has a well-defined finite coefficient, and we can tune the rest of the 
amplitude on it, in unitary gauge this coefficient is indeterminate, 
possibly divergent: we must then add another counterterm 
$\delta g_{0\!A} h_0 \left(\partial^\mu A_0^\nu\right)^2$ 
to absorb the divergence.

We have now {\it two counterterms} and we need {\it two renormalization conditions} 
to fix the arbitrariness. The only Dyson subtraction 
(which means imposing gauge invariance) is not enough anymore. 
The result in Eq.~(\ref{eq:amplitudewrong}) is still arbitrary, 
and allows the addition of whatever gauge invariant 
$k_2^\mu k_1^\nu - k_1 \!\cdot k_2\, g^{\mu\nu}$ term. 
The other condition could be, for example, 
the requirement of the validity of the 
equivalence theorem~\cite{equivalenceth} in the limit $m_W \to 0$, 
or the invariance of the amplitude in both `t~Hooft-Feynman and unitary gauge:
both conditions fix the value of the amplitude in Eq.~(\ref{eq:amplitudewrong}) 
to the standard result in  Eq.~(\ref{eq:amplitudecorrect}). 
The indeterminacy is resolved; we recover also the independence of the amplitude on gauge choice and regulator choice.

\section{Renormalizable gauges}
\label{sec:4}
The main difference between renormalizable $R_\xi$ gauges and 
the unitary gauge is in  the ultraviolet behaviour of the propagator
\begin{eqnarray}
 i\Delta^{\mu\nu} &=& \frac{1}{q^2 - m_W^2}\left(g^{\mu\nu} - \left(1 - \xi\right) \frac{q^\mu q^\nu}{q^2 - \xi m_W^2}\right) \label{eq:propagatorW} \nonumber\\
&=& \frac{g^{\mu\nu} - \frac{q^\mu q^\nu}{m_W^2}}{q^2 - m_W^2} + \frac{q^\mu q^\nu}{m_W^2} \frac{1}{q^2 - \xi m_W^2}
\end{eqnarray}
For finite $\xi $, the propagator is $O\left(q^{-2}\right)$, 
and leads to a logarithmic divergent integral, the 
same as in `t~Hooft-Feynman gauge; 
on the other hand, for \mbox{$\xi = \infty$} (unitary gauge) the propagator is $O\left(q^0\right)$, 
leading to a highly-divergent amplitude. 
If we use a gauge-invariant regulator, we could simply handle 
all high divergences and show that all $\xi$-dependent terms in the 
amplitude vanish, leading to the same result as in the unitary 
gauge \cite{Marciano:2011}. 
In other words, since gauge-invariant regularizations 
give meaning to the integrals, we can carry the limit 
\mbox{$\xi = \infty$} under the integral.

On the contrary, in gauge-breaking regularization, divergent integrals 
have ambiguities, and we have to handle limits with care. 
Let us study the divergent behaviour in $R_\xi$ gauges. 
The integrals remain logarithmic divergent, 
so we can perform usual Feynman parametrization and shift 
the integration variables.
We obtain the expression
\begin{eqnarray}
{\cal M}^{\mu\nu}_{\xi<\infty} &=& \frac{e^2 g}{i\left(2\pi\right)^4 m_W} \left(6m_W^2 + m_H^2\right) 5! \int_0^1 \prod_{i=1}^6 dx_i \,\delta\!\left(1 - \sum_{i=1}^6 x_i\right)  \nonumber\\
&{}& \cdot \int d^4 l\,\frac{l^6}{\left(l^2 - M^2 + i\epsilon\right)^6} \left(4 l^\mu l^\nu - l^2 g^{\mu\nu}\right) + \rm{finite\,\, integrals}
\label{eq:amplitudexi}
\end{eqnarray}
with $M^2 = m_W^2 \left(1 + \left(\xi - 1\right)\left(x_4 + x_5 + x_6\right)\right) - m_H^2 \left(x_3 + x_6\right)\left(x_1 + x_4\right)$. 
Despite of the complication of Feynman parameters, 
we have the same form as in Eq.~(\ref{eq:amplitudefeynman}): 
the integral can be treated as an indeterminate 
tensor $I g^{\mu\nu}$, while the coefficient of $k_2^\mu k_1^\nu$ 
is finite and uniquely determined. 
As in the previous section, by Dyson subtraction we can 
tune the indeterminate constant and recover the correct 
gauge invariant amplitude in Eq.~(\ref{eq:amplitudecorrect}).

In the unitary gauge, the behaviour is completely different: 
the divergent integral in Eq.~(\ref{eq:amplitudexi}) should vanish 
in the  $\xi \to \infty$ limit, so the divergences in unitary 
gauge must arise from a combination of highly-divergent terms. 
Indeed, let us consider again the calculation of Ref.~\cite{Gastmans:2011wh}. 
Putting external momenta on-shell, 
Gastmans {\etal} show that high divergences reduce 
 to  $O\left(k^2\right)$ terms, and then get simplified  
to the logarithmic divergent term in Eq.~(\ref{eq:aprimo}) 
and in a quadratically divergent term which, 
with an appropriate choice of loop momentum, 
vanishes for tensor 
invariance~\footnote{An integral of 
the form $\int d^4 l \frac{l^\alpha}{{\rm even\,polinomial\,in}\,l}$ 
must vanish because we have no 1-index constant tensor 
it can be proportional to. Obviously, it would vanish also 
in symmetric integration.}.

To summarize, the indeterminate behaviour of $k_2^\mu k_1^\nu$ 
in the unitary gauge arises from highly divergent terms which do not 
appear in $R_\xi$ gauges. The limit $\xi \to \infty$ cannot be 
taken under the integral because integrals are ill-defined, 
and integrand functions are not measurable. 
The amplitude is arbitrary only in the unitary gauge, 
and divergences apparently disappear only with a particular 
choice of loop momentum.

\section{Conclusions}
\label{sec:5}
We have analyzed the computation of the amplitude $H \to \gamma\gamma$ by Gastmans {\etal} \cite{Gastmans:2011ks,Gastmans:2011wh}, to understand why 
it turns out to be different from the standard result in Eq.~(\ref{eq:amplitudecorrect}). 
Integrals of the form
\begin{equation}
 I_{\mu\nu} = \int d^4 l \, \frac{g_{\mu\nu} l^2 - 4 l_\mu l_\nu}{\left(l^2 - M^2+i \epsilon \right)^3} 
\label{eq:anotherintegral4D}
\end{equation}
are not well defined. We have provided some explicit examples  
within cutoff regularization, obtaining different values by varying integration boundaries.

In `t~Hooft-Feynman gauge and in a cutoff regularization scheme,  see Eq.~(\ref{eq:amplitudefeynman}), we  obtain
\begin{equation}
{\cal M}^{\mu\nu}_{\xi=1} = \frac{e^2g}{(4\pi)^2m_W} \left[ -k_2^\mu k_1^\nu\left(2 + 3 \tau + 3\tau\left(2-\tau\right)f\left(\tau\right)\right) + I g^{\mu\nu} \right]
\label{eq:anotheramplitudefeynman}
\end{equation}
where $I$ is a constant which depends on the boundary shape. This makes it indeterminate 
as there is no {\it physical prescription} on the choice of the integration boundary shape.
On the other hand, the use of a gauge invariant regularization scheme provides automatically 
the recipe on how to evaluate the integrals.

Since the term $k_2^\mu k_1^\nu$, in Eq.~(\ref{eq:anotherintegral4D}), has {\it only one} finite unambiguos coefficient, we are able to solve the indeterminacy by imposing gauge invariance at the end of the calculation. 

However we have shown that, in the unitary gauge, 
both the coefficients  $k_2^\mu k_1^\nu$ and $g^{\mu\nu}$ are indeterminate in the sense of integration boundary shape dependency. 
Imposing only one renormalization condition (like imposing gauge invariance by Dyson subtraction) 
{\it is not enough anymore}.  Given the equivalence of $R_\xi$ gauges with unitary gauge as $\xi\to \infty$, 
the problem we discuss is likely related to the exchange of this limit
with integral sign for  not Riemann-summable functions: the coefficient of $k_2^\mu k_1^\nu$ arises from highly divergent terms which do not appear at finite values of $\xi$. 

Gastmans {\etal}'s expression in Eq.~(\ref{eq:amplitudewrong}) 
is still ambiguous upon Dyson subtraction, and allows the addition of whatever 
term of the form $k_2^\mu k_1^\nu - k_1 \!\cdot k_2\, g^{\mu\nu}$. 
This arbitrariness can be fixed by requiring the validity of the 
equivalence theorem, or by imposing the equality of amplitudes in unitary 
and `t~Hooft-Feynman gauges. In other words, we are able to add terms 
to Eq.~(\ref{eq:amplitudewrong}) in order to match the standard 
result Eq.~(\ref{eq:amplitudecorrect}).

The combination of unitary gauge with a cutoff regularization scheme simply turns out to be non-predictive. 

\appendix
\section{Appendix}
\label{sec:appendix}
For the sake of simplicity in the following  we consider a two-dimensional version of $I_{\mu\nu}$ in Eq.~(\ref{eq:integral4DWick}) 
\begin{equation}
 I_{\mu\nu} = i \int d_2 l \, 
\frac{\delta_{\mu\nu} l^2 - 2 l_\mu l_\nu}{\left(l^2 + 1\right)^2} 
\label{eq:integralD2}
\end{equation}

This version of $I_{\mu\nu}$ has the same properties of its four-dimensional counterpart, namely:
$i)$~the integral is superficially divergent as a logarithm, $ii)$~it is identically zero for symmetric integration domains, $iii)$~the integrand function has no definite sign. 
The conclusions we will draw from the following calculations in two dimensions remain unaltered in four dimensions: here we just  avoided superfluous technical complications.

As we did before, let us start by computing the $I_{11}$ term. We realize  that $I_{11}$ can be mapped into an entry of the $I_{12}$ kind upon a rotation by 45$^{o}$ of $l_{1}l_{2}$ axes. 
$I_{11}=I_{12}$ only if the integration domain is rotated accordingly.
Since the calculations turn out to be simpler using the $\{12\}$ entry, 
we will make our observations on this case only
\begin{equation}
 I_{12} = i \int d_2 l \frac{-2l_1 l_2}{\left(1 + l^2\right)^2} = i \int d_2 l \, F_{12}
\label{eq:integral2D}
\end{equation}
At any rate we remark  that the domains of integrations will be chosen in such a way that eventually all off-diagonal $I_{\mu\nu}$ entries will vanish as to recover eventually the $\delta_{\mu\nu}$ tensor structure.

\begin{figure}
\centering%
\includegraphics[width=7cm]{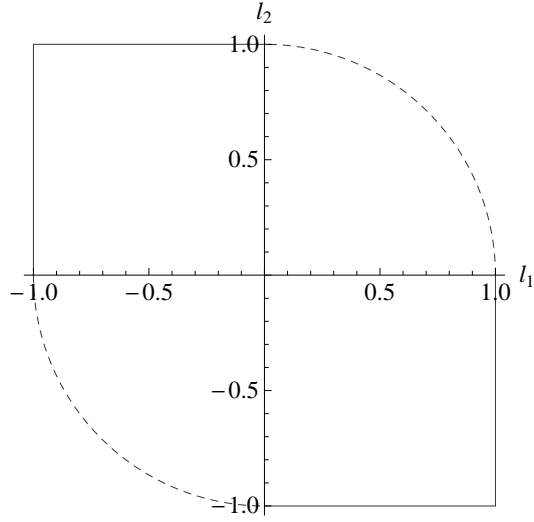}%
\caption{We integrate $F_{12} = \frac{-2l_1 l_2}{\left(1 + l^2\right)^2}$ over a mixed boundary. We bound the domain with a circle when $F_{12}<0$, and with a square otherwise. Since we broke spherical symmetry, the integral does not vanish.}%
\label{fig:circsquare}%
\end{figure}

The integrand in Eq.~(\ref{eq:integral2D}) is negative when $l_1 l_2 > 0$ (I and III quadrant), and positive otherwise. In the former case, we bound a domain with two quarters of  a circumference of radius $\Lambda$; in the latter case we use a square with edge $\Lambda$ (\figurename{} \ref{fig:circsquare}).
We have
\begin{eqnarray}
 I_{12} &=& - 4 i \int_0^\Lambda d l \, \frac{l^3}{\left(1 + l^2\right)^2} \, \int_0^{\pi/2} d \theta \sin\theta \cos\theta + 4 i \int_{\left[0,\Lambda\right]\times\left[0,\Lambda\right]}\!\!\!\!\!\!\!\!\!\!\!\!\!\!\!\!\!\!dl_1 dl_2 \,\,\, \frac{l_1 \, l_2}{\left(1+l_1^2 + l_2^2\right)^2} \nonumber \\
 &=& i \left( \frac{\Lambda^2}{1 + \Lambda^2} + 
   \ln\frac{1 + \Lambda^2}{1 + 2 \Lambda^2}\right) = i \left(1 + \ln\frac12\right) + O\left(\frac1{\Lambda^2}\right)
\label{eq:integral2Dfinite}
\end{eqnarray}
Again we get a finite non-zero value. The leading divergences are the same in each quadrant, whereas the finite part is boundary-dependent, so that the sum does not vanish. 

More in general, we can slice $\R^2$ in a countable set of 
bounded regions, in order to reduce the integral over the whole $\R^2$ to a countable sum of finite integrals, i.e., to a series. We can thus use the Riemann rearrangement theorem \cite{Apostol:2000ap} to obtain whatever finite value or logarithmic divergence.

\begin{figure}
\centering%
\subfigure[\label{fig:riemann1}]{\includegraphics[width=4.3cm]{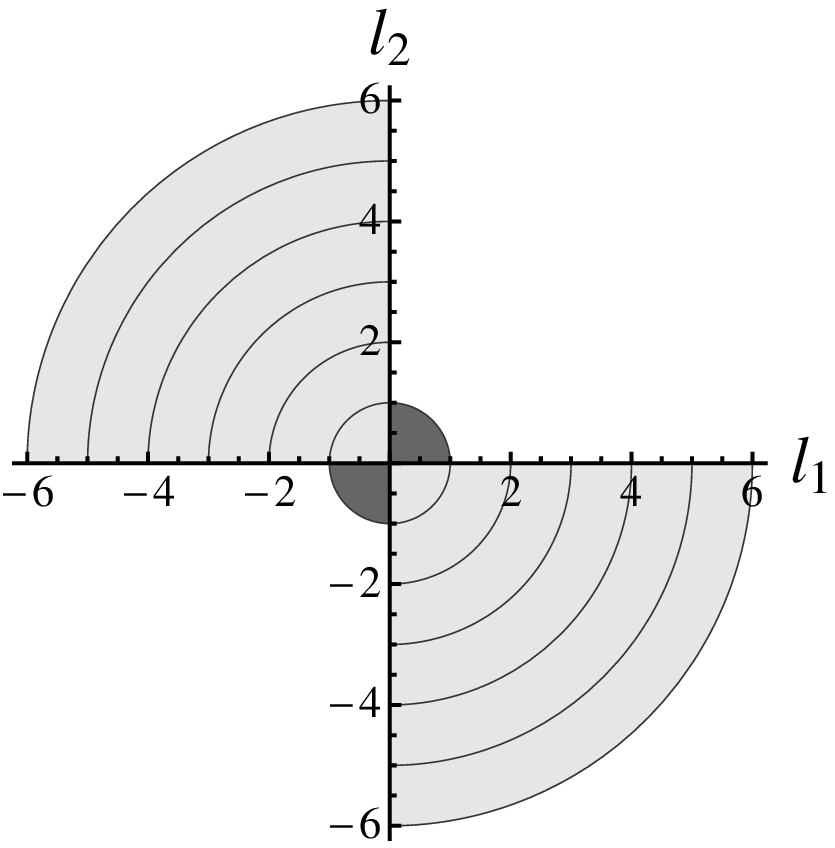}}\quad
\subfigure[\label{fig:riemann2}]{\includegraphics[width=4.3cm]{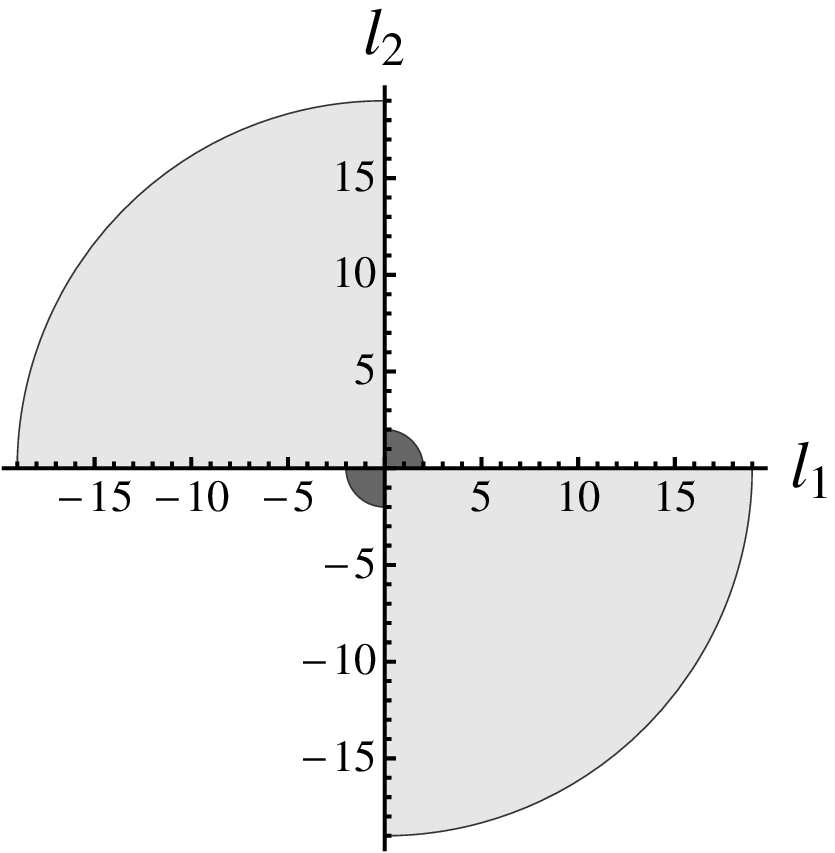}}\quad
\subfigure[\label{fig:riemann3}]{\includegraphics[width=4.3cm]{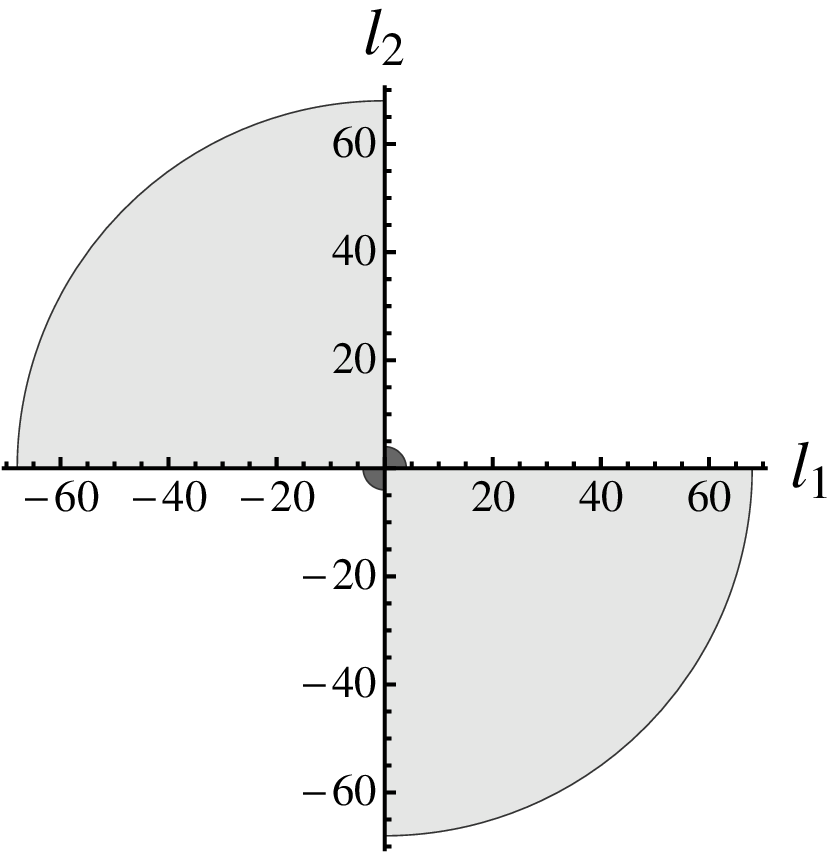}}%
\caption{Riemann rearrangement. Light gray regions have positive integral $p_k$, dark gray regions have negative integrals $n_k$. The absolute value of each region is bounded by $M \approx 0.62$. We can obtain a divergent sum following this algorithm: (a) we sum first positive terms $p_0 + \dots + p_{N_1}$, until we exceed $1 + M$, then we can subtract $n_0$ still exceeding $1$; (b) we continue adding positive terms until we exceed $2+M$, then we can subtract $n_1$ still exceeding $2$; (c) and so on. We see that the negative region becomes smaller and smaller than the positive region, so that cannot cancel the logarithmic divergence.}%
\label{fig:riemann}%
\end{figure}

For example, let us consider all the concentric circumferences 
with integer radius thus slicing $\R^2$ into annuli: the integral of $F_{12}$ over each annulus vanishes by circular symmetry. Therefore we slice each annulus into a positive region $P_k$ where $F_{12}>0$, and a negative region $N_k$ where $F_{12}<0$ (\figurename{} \ref{fig:riemann}). We therefore have
\begin{eqnarray}
 p_k &=& \int_{P_k} d_2 l\, F_{12} = 4 \int_0^{\pi/2} d\theta \sin\theta \cos\theta \, \int_{k}^{k+1} d l\,\frac{l^3}{\left(1 + l^2\right)^2}\nonumber\\
&=& \frac1{k^2 + 2k + 2} - \frac1{k^2 + 1} + \log\frac{k^2 + 2k +2}{k^2 +1} \nonumber\\
n_k &=& \int_{N_k} d_2 l\, F_{12} = -p_k
\end{eqnarray}
The $p_k$ form a bounded sequence of positive terms converging to 0. We can find that the greatest term of the sequence is $p_1 = M \approx 0.62$. Specularly, the $n_k$ form a sequence of negative terms converging to zero, bounded by $n_1 = -M \approx -0.62$. If we the the union of all $P_k$ and $N_k$, 
we recover the whole $\R^2$, therefore if we sum all $p_k$ and $n_k$ we recover the whole integral. Since $\sum_k p_k$ and $\sum_k n_k$ both diverge separately, we must specify the correct ordering of terms. We start by adding the first positive terms $p_k$ until we exceed $1+M$, and then add the first negative term $n_0$. Since all $0 > n_k > -M$, we have still
\begin{equation}
p_0 + p_1 + \dots + p_{N_1} - \left|n_0\right| > 1
\end{equation}
We can continue adding positive terms until we exceed $2 + M$, and then add $n_1$, and so on. The resulting sum covers all $P_k$ and $N_k$ regions. The series diverges, and so does the integral.

\begin{figure}
\centering%
\includegraphics[width=7cm]{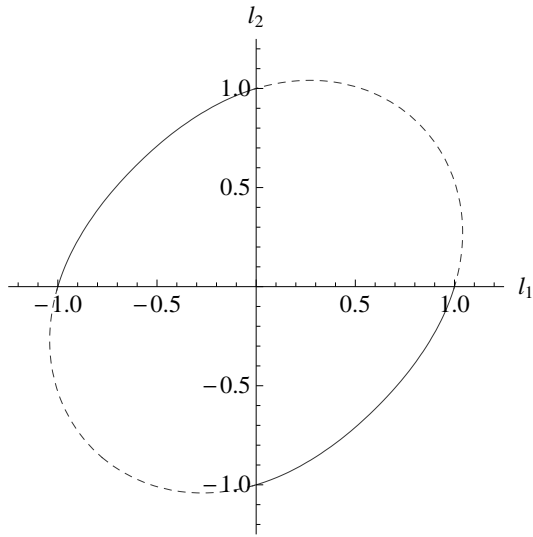}%
\caption{We integrate $F_{12} = \frac{-2l_1 l_2}{\left(1 + l^2\right)^2}$ regulating the function with a smooth cutoff $\Lambda$. In the graphic we show a deformed cutoff $\Lambda\left(\theta\right)=\Lambda e^{\epsilon \sin 2\theta}$. The curve is filled in the region where $F_{12}>0$, dashed otherwise. Since we broke spherical symmetry, the integral does not vanish.}%
\label{fig:schwinger}%
\end{figure}
One might wonder whether we have the same behavior with a smooth cutoff. 
We  calculate Eq.~(\ref{eq:integral2D}) 
with a Schwinger regulator \cite{Schwinger} 
\begin{eqnarray}
 I_{12} &=& i \int d_2 l \left(-2l_1 l_2\right) \int_{\frac1{\Lambda^2}}^\infty ds\,s\, e^{-s\left(1 + l^2\right)}\nonumber \\
&=& -i\, \Gamma\!\left(0,\frac1{\Lambda^2}\right) \int_0^{2\pi} d \theta\,\sin\theta \cos\theta = 0
\label{eq:integralschwinger}
\end{eqnarray}
where $\Gamma\!\left(a,b\right)$ is the incomplete Gamma function \cite{Abramowitz:1972}. Again, the angular part of the integral vanishes, 
and so we do not care about the logarithmic divergence in the radial part. 
However if we deform the  cutoff giving an angular dependency to it, e.g. $\Lambda \to \Lambda\left(\theta\right)=\Lambda \exp\left(\epsilon \sin 2\theta\right)$ (\figurename{} \ref{fig:schwinger}), we obtain
\begin{eqnarray}
 I_{12} &=& i \int d_2 l \left(-2l_1 l_2\right) \int_{\frac1{\Lambda^2\left(\theta\right)}}^\infty ds\,s\, e^{-s\left(1 + l^2\right)}\nonumber \\
 &=& -i \int_0^{2\pi} d \theta\,\sin\theta \cos\theta \,\Gamma\!\left(0,\frac1{\Lambda^2\!\left(\theta\right)}\right) = -i \pi \epsilon + O\left(\frac1{\Lambda^2}\right)
\label{eq:integralschwingerasymm}
\end{eqnarray}
Also in this case, the value of the integral depends on the shape of the cutoff function, no matter if smooth or sharp.

\end{document}